\documentclass{aa}

\usepackage{amsmath}
\usepackage{graphicx}
\usepackage{graphics}
\usepackage{epsfig}
\usepackage{txfonts}
\usepackage{flushend}
\begin{document}

\title{Steep extinction towards GRB\,140506A reconciled from host galaxy observations: Evidence that steep reddening laws are local\thanks{Based on observations carried out under programme IDs 095.D-0043(A, C) and 095.A-0045(A) with the X-shooter spectrograph and the FOcal Reducer and low dispersion Spectrograph 2 (FORS2) installed at the Cassegrain Very Large Telescope (VLT), Unit 2 – Kueyen and Unit 1 - Antu, respectively, operated by the European Southern Observatory (ESO) on Cerro Paranal, Chile.}}

\author{K.~E.~Heintz\inst{1,2},
J.~P.~U.~Fynbo\inst{2},
P.~Jakobsson\inst{1},
T.~Kr\"{u}hler\inst{3},
L.~Christensen\inst{2},
D.~Watson\inst{2},
C.~Ledoux\inst{4},
P.~Noterdaeme\inst{5},
D.~A.~Perley\inst{6,2},
H.~Rhodin\inst{2},
J.~Selsing\inst{2},
S.~Schulze\inst{7},
N.~R.~Tanvir\inst{8},
P.~M\o ller\inst{9},
P.~Goldoni\inst{10},
D.~Xu\inst{11},
B.~Milvang-Jensen\inst{2}
}

\institute{Centre for Astrophysics and Cosmology, Science Institute, University of Iceland, Dunhagi 5, 107 Reykjav\'ik, Iceland
\and
Dark Cosmology Centre, Niels Bohr Institute, University of Copenhagen, Juliane Maries Vej 30, 2100 Copenhagen Ø, Denmark
\email{keh14@hi.is}
\and
Max-Planck-Institut f\"{u}r extraterrestrische Physik, Giessenbachstra\ss e, 85748 Garching, Germany
\and
European Southern Observatory, Alonso de C\'ordova 3107, Vitacura, Casilla 19001, Santiago 19, Chile
\and
Institut d'Astrophysique de Paris, CNRS-UPMC, UMR7095, 98bis bd Arago, 75014 Paris, France
\and
Astrophysics Research Institute, Liverpool John Moores University, IC2, Liverpool Science Park, 146 Brownlow Hill, Liverpool L3 5RF, UK
\and
Department of Particle Physics and Astrophysics, Weizmann Institute of Science, Rehovot 7610001, Israel
\and
University of Leicester, Department of Physics and Astronomy, University Road, Leicester, LE1 7RH, United Kingdom
\and
European Southern Observatory, Karl-Schwarzschildstrasse 2, D-85748 Garching bei M\"unchen, Germany
\and
APC, Univ. Paris Diderot, CNRS/IN2P3, CEA/Irfu, Obs. de Paris, Sorbonne Paris Cit\'e, France
\and
CAS Key Laboratory of Space Astronomy and Technology, National Astronomical Observatories, Chinese Academy of Sciences, Beijing 100012, China
}

\titlerunning{Reconciling the steep extinction towards GRB\,140506A}
\authorrunning{Heintz et al.}

\date{Received 2017; accepted, 2017}

\abstract{
We present the spectroscopic and photometric late-time follow-up of the host
galaxy of the long-duration \textit{Swift} $\gamma$-ray burst
GRB\,140506A at $z=0.889$. The optical and near-infrared afterglow of this GRB 
had a peculiar spectral energy distribution (SED) with a strong flux-drop at 
8000\,\AA~(4000\,\AA~rest-frame) suggesting an unusually steep
extinction curve. By analysing the contribution and physical properties of the host galaxy, we 
here aim at providing additional information on the properties and origin of this 
steep, non-standard extinction. We find that the strong flux-drop in the GRB
afterglow spectrum at $< 8000$\,\AA~and rise at $< 4000$\,\AA~(observers frame)
is well explained by the combination of a steep extinction curve along the GRB
line of sight and contamination by the host galaxy light at short wavelengths
so that the scenario with an extreme 2175\,\AA~extinction bump can be excluded.
We localise the GRB to be at a projected distance of approximately 4 kpc from
the centre of the host galaxy. Based on
emission-line diagnostics of the four detected nebular lines, H$\alpha$, H$\beta$,
[O\,\textsc{ii}] and  [O\,\textsc{iii}], we find the host to be a modestly 
star forming (SFR =
$1.34\pm 0.04~M_{\odot}$ yr$^{-1}$) and relatively metal poor ($Z=0.35^{+0.15}_{-0.11}~Z_{\odot}$) 
galaxy with a large dust content, characterised by a measured
visual attenuation of $A_V=1.74\pm 0.41$ mag. We compare the host
to other GRB hosts at similar redshifts and find that it is unexceptional in
all its physical properties. We model the extinction curve of the
host-corrected afterglow and show that the standard dust properties causing the
reddening seen in the Local Group are inadequate in describing the steep drop.
We thus conclude that the steep extinction curve seen in the afterglow towards
the GRB is of exotic origin and is sightline-dependent only, further confirming
that this type of reddening is present only at very local scales and that it is
solely a consequence of the circumburst environment.}

\maketitle

\section{Introduction}     
\label{sec:introduction}

Gamma-ray bursts (GRBs) have proven to be valuable tools to probe the
interstellar medium (ISM), chemical enrichment and dust content of star-forming
galaxies both in the local Universe and out to high redshifts
\citep[e.g.][]{Jakobsson04,Fynbo06,Fynbo09,Prochaska07,Li08,Ledoux09,Gehrels09,Schady12,DeCia13,Thoene13,Sparre14,Hartoog15}.
Specifically, the optical and near-infrared extinction curves seen toward GRB
afterglows reveal that they typically can be well-described by similar
prescriptions to that of the Milky Way (MW) and the Small and Large Magellanic
Clouds (SMC and LMC) \citep[see e.g.][]{Schady10,Covino13,Japelj15}. In rare cases,
however, the afterglows appear to show flat (also known as grey) extinction
curves \citep{Savaglio04,Perley08,Friis15} or at the opposite extreme, very steep extinction curves. In
the latter case, a prime example is seen towards the afterglow of GRB\,140506A
\citep{Fynbo14} but also reported in the spectral energy distribution (SED) of the afterglow of GRB\,070318 \citep{Fynbo09,Watson09} and bluewards of the observed 2175 \AA~bump in GRB\,080605A \citep{Zafar12,Kruehler12}. 

\begin{figure}
	\centering
	\epsfig{file=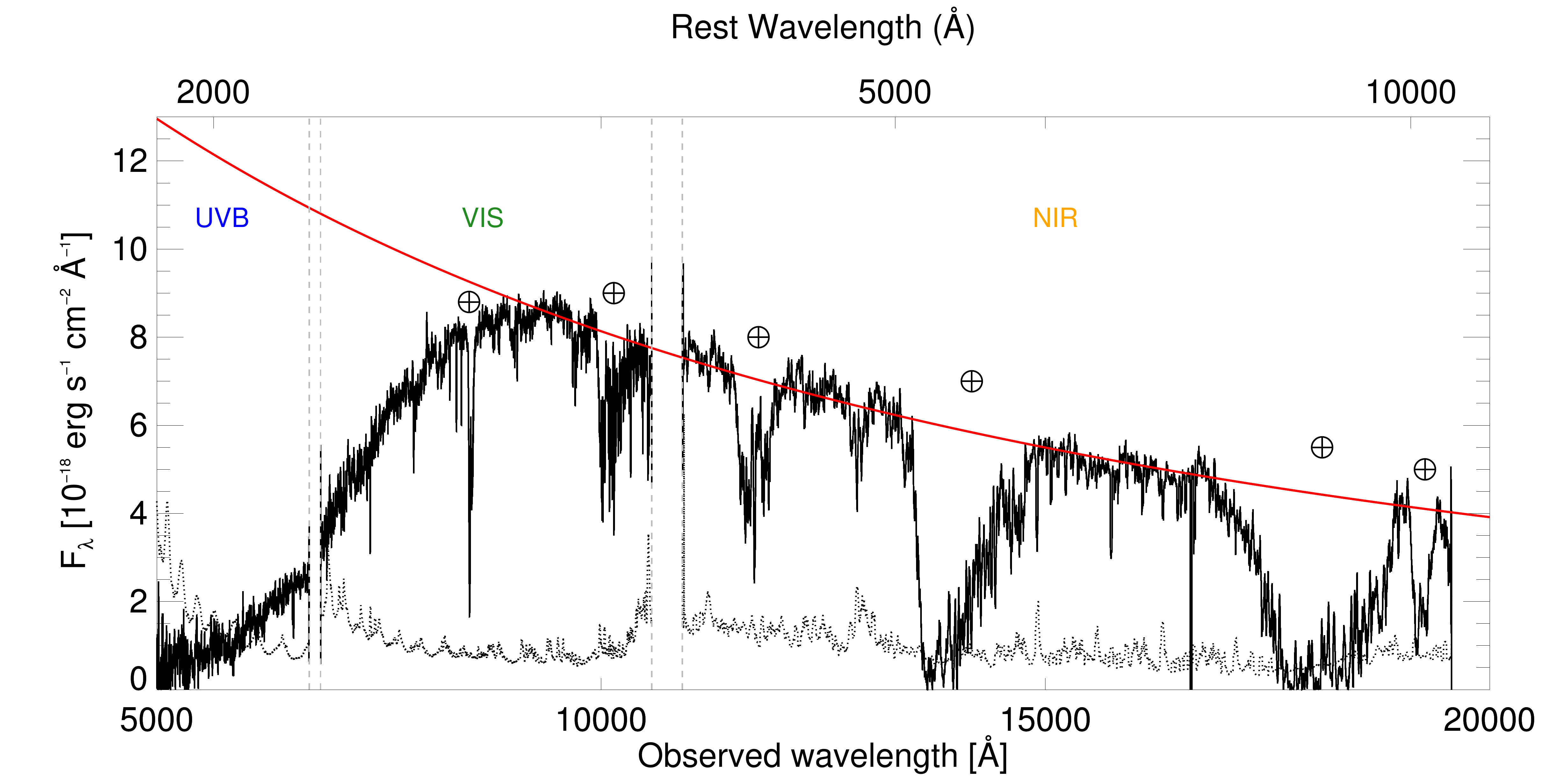,width=9cm,height=5.5cm}
	\caption{X-shooter spectrum of the first epoch afterglow \citep[identical to the top panel in Fig.~4 of][]{Fynbo14}. Regions affected by telluric absorption are marked by the corresponding symbol. The error spectrum is plotted as the dotted line. Overplotted is the best-fit power law following $F_{\lambda}\propto \lambda^{-\beta}$, fitted to the afterglow continuum at $> 8000$\,\AA~shown by the red solid line. \label{fig:GRBOA}}
\end{figure}

This paper reports on the spectroscopic and photometric late-time follow-up of
the host galaxy of GRB\,140506A at $z=0.889$. This GRB was first detected by
\textit{Swift} on 2014 May 6, 21:07:36 UT \citep{Gompertz14}. At 8.8 and 33
hours post-burst, \cite{Fynbo14} acquired spectra with the X-shooter
spectrograph \citep{Vernet11} which, together with extensive imaging from the
Gamma-Ray burst Optical and Near-infrared Detector \citep[GROND;][]{Greiner08},
revealed a peculiar optical and near-infrared afterglow and the first detections of He~\textsc{i}* and Balmer line absorption in GRBs. In contrast to other GRB
afterglow SEDs, which typically can be modelled with a simple power-law with
weak SMC-like extinction, a strong, gradual flux drop is seen below 8000 \AA~(4000 \AA~rest-frame, see Fig.~\ref{fig:GRBOA}) that can not be modelled by any of the extinction curves known from the Local Group. By analysing the host we here aim to better constrain the physical origin of this steep extinction curve.

The paper is structured as follows. In Section~\ref{sec:obs} we describe our
observations and in Sects.~\ref{sec:res} and \ref{sec:disc} we describe how we constrained the properties of the host galaxy with special emphasis on the implications for the interpretation
of the extinction derived from the afterglow spectroscopy.
In Section~\ref{sec:conc} we summarise and conclude. Throughout the paper we assume a
standard flat $\Lambda$CDM cosmology with $H_0=67.3$ km s$^{-1}$ Mpc$^{-1}$,
$\Omega_{\mathrm{m}}=0.315$ and $\Omega_{\Lambda}=0.685$ \citep{Planck14}.
Unless otherwise stated, all magnitudes are given in the AB \citep{Oke74}
magnitude system.

\section{Observations and data reduction} \label{sec:obs}

Based on the peculiar (and temporally varying) SED of the optical and near-infrared afterglow of GRB\,140506A, acquired at two epochs at $\Delta$t = 8.8 hr and $\Delta$t = 33 hr after the burst, respectively, we decided to observe the host galaxy of the GRB to examine the nature of this in more detail than what can be extracted from the afterglow spectra. 

\begin{table*}[!htbp]
	\centering
	\begin{minipage}{0.85\textwidth}
		\centering
		\caption{Photometry of the host galaxy from FORS2 and \textit{Spitzer}/IRAC imaging. \label{tab:photdata}}
		\begin{tabular}{cccccccc}
			\noalign{\smallskip} \hline \hline \noalign{\smallskip}
			Date ($\Delta$t)\tablefootmark{a} & Filter  & Eff. wavelength & Exp. time & Avg. airmass & Seeing\tablefootmark{b} & Mag$_{\mathrm{AB}}$\tablefootmark{c} \\
			\noalign{\smallskip}\hline \noalign{\smallskip}
			2015-04-29 (357.48 d) & $u_{\mathrm{HIGH}}$ & 3610 \AA & $3\times500$ s & 1.166 & $0\farcs62$ & $25.60\pm 0.30$ \\
			2015-04-29 (357.50 d) & $g_{\mathrm{HIGH}}$ & 4700 \AA & $3\times300$ s & 1.171 & $0\farcs60$ & $25.47\pm 0.21$ \\
			2015-06-16 (405.24 d) & $R_{\mathrm{SPECIAL}}$ & 6550 \AA & $5\times200$ s & 1.323 & $0\farcs70$ & $25.16\pm 0.19$ \\
			2015-06-16 (405.26 d) & $I_{\mathrm{BESS}}$ & 7680 \AA & $5\times200$ s & 1.280 & $0\farcs67$ & $24.46\pm 0.20$  \\
			2015-07-04 (423.12 d) & IRAC 1 & 3.6 $\mu$ m & 2400 s & $\cdots$ & $\cdots$ & $23.88 \pm 0.30 $ \\ 
			\noalign{\smallskip} \hline \noalign{\smallskip}
		\end{tabular}
		\centering
		\tablefoot{\\
			\tablefoottext{a}{The date of observations are listed with the number of days from the trigger in brackets. The time is calculated from the start of observations to the time of trigger.}\\
			\tablefoottext{b}{Measured from the full-width-at-half-maximum of non-saturated stellar sources on the stacked images.}\\
			\tablefoottext{c}{Magnitudes reported here have been corrected for the Galactic foreground reddening and will be used throughout our analysis.}
		}
		\end{minipage}
\end{table*}

\subsection{FORS2 imaging}

We acquired late-time photometry
with the FOcal Reducer and low dispersion Spectrograph 2 \citep[FORS2;][]{Appenzeller98} instrument at
the ESO/VLT on 2015 April 29 and June 16. 
We obtained images of the field of
GRB\,140506A using the $u_{\mathrm{HIGH}}$, $g_{\mathrm{HIGH}}$,
$R_{\mathrm{SPECIAL}}$ and $I_{\mathrm{BESS}}$ broadband filters (henceforth,
these will be referred to as simply the $u$-, $g$-, $R$- and $I$-band,
respectively). In Table~\ref{tab:photdata} a log of the FORS2 observations is
provided together with the measurements of the magnitude of the host galaxy in each filter. 

The photometric calibration of the observations in the  $u$- and $g$-bands was 
done using zeropoints measured from reference stars observed at similar airmasses on the nights of the observations of the host galaxy. For the $R$- and $I$-bands we use the zeropoints reported in the ESO archive for the night of observation. All images were taken following a dithering pattern and were bias-subtracted and
flat-fielded using standard \texttt{IRAF} \citep{Tody93} routines. The images acquired in
April were taken under excellent conditions with a median seeing of $0\farcs61$,
where for the images obtained in June we report a median seeing of $0\farcs69$.
To measure the magnitude of the host we performed aperture photometry of the source in the combined frames of each filter using standard MIDAS routines. 

\subsection{Spitzer/IRAC observations}

The position of GRB 140506A was observed by Spitzer/IRAC in the 3.6 $\mu$m filter as part of the extended \textit{Swift/Spitzer} GRB Host Galaxy Survey \citep[SHOALS;][GO program 11116]{Perley16a} on 2015 July 4, with a total integration time of 2400 seconds.  We downloaded the reduced imaging (PBCD) from the \textit{Spitzer} Heritage Archive.  The area close to the GRB host galaxy is affected by a weak pulldown artifact from a nearby star; we fixed this manually by adding a constant term to the affected columns. The host galaxy is faint but clearly detected in this band, and there is no strong blending from any nearby sources.  We measure photometry of the galaxy again using standard aperture photometry techniques using the zeropoints in the IRAC handbook for the absolute calibration. 

\subsection{X-shooter spectroscopy}

To supplement the FORS2 and \textit{Spitzer} imaging we acquired additional spectra centred on the host galaxy using the X-shooter spectrograph mounted at the ESO/VLT, covering a wavelength range of 3000--24\,800\,\AA~(3000--5600, 5500--10\,200 and 10\,200--24\,800\,\AA~for the UVB, VIS and NIR
arm, respectively). The follow-up observations of this host were spread out
over three nights, on 2015 May 15, June 15 and September 30. The reason for this
was that we realised that the heliocentric velocity shift had placed the
H$\alpha$ line on top of a bright sky emission line in May and June. In
September the velocity caused by the Earth's orbit was optimal for separating
the H$\alpha$ emission line away from the skyline. 

\begin{table}[!htbp]
	\centering
	\begin{minipage}{\columnwidth}
		\centering
		\caption{Spectroscopic observations of the host galaxy with X-shooter. \label{tab:specdata}}
		\begin{tabular}{cccc}
			\noalign{\smallskip} \hline \hline \noalign{\smallskip}
			Date ($\Delta$t)\tablefootmark{a} & Exp. time  & Avg. airmass & Seeing\tablefootmark{b} \\
			\noalign{\smallskip}\hline \noalign{\smallskip}
			2015-05-15 (373.47 d) & $4\times600$ s & 1.198 &$0\farcs72$ \\
			2015-06-15 (404.31 d) & $4\times600$ s & 1.184 & $0\farcs90$ \\
			2015-06-15 (404.36 d) & $4\times600$ s & 1.173 & $0\farcs98$ \\
			2015-06-15 (404.41 d) & $5\times600$ s & 1.268 & $0\farcs84$ \\
			2015-09-30 (511.11 d) & $4\times600$ s & 1.233 & $1\farcs24$ \\
			2015-09-30 (511.15 d) & $4\times600$ s & 1.332 & $1\farcs18$ \\
			2015-09-30 (511.19 d) & $4\times600$ s & 1.492 & $0\farcs81$\\
			\noalign{\smallskip} \hline \noalign{\smallskip}
		\end{tabular}
		\centering
	\end{minipage}
	\tablefoot{\\
		\tablefoottext{a}{Same notation as in Table~\ref{tab:photdata}.}\\
		\tablefoottext{b}{The seeing is measured from the telescope guide probe.}
	}
\end{table}

The observations were carried out more than a year after the GRB and 
we can hence safely assume that there is no significant contribution from 
the afterglow to the measured fluxes. Extrapolating the temporal flux decay from the GROND afterglow observations suggests that the afterglow contributes less than 5\% of the host galaxy flux at this point.
The observations were all carried out under good conditions with a
seeing between $\sim 0\farcs7 - 1\farcs2$ at airmasses between 1.1 and 1.5. The
observations in May were carried out in a $4\times 600$ s observing block (OB),
while in June and September we acquired three executions of $4\times 600$ s
(except for the last OB in June where we acquired $5\times 600$ s), following
an ABBA nodding pattern. The slit was aligned with the parallactic angle, and
for the multiple executions in June and September we re-aligned the slit
between the three executions of the OBs. The slit widths were $1\farcs 0$,
$0\farcs 9$ and $0\farcs 9$ in the UVB, VIS and NIR arm, respectively, for all
observations. The details for the X-shooter observations are all listed in
Table~\ref{tab:specdata}. For this given setup the nominal instrumental
resolution in the respective arms are $\mathcal{R}_{\mathrm{UVB}}=5100$,
$\mathcal{R}_{\mathrm{VIS}}=8800$ and
$\mathcal{R}_{\mathrm{NIR}}=5100$\footnote{\url{https://www.eso.org/sci/facilities/paranal/instruments/xshooter/inst.html}}
($\approx$ 59, 34 and 59 km s$^{-1}$, respectively). From a set of atmospheric 
emission lines in the VIS and NIR arms we measure resolutions of
$\mathcal{R}_{\mathrm{VIS}}=9096$ ($\approx$ 33 km s$^{-1}$) and
$\mathcal{R}_{\mathrm{NIR}}=6977$ ($\approx$ 43 km s$^{-1}$) at
full-width-at-half-maximum (FWHM). These values were determined from sky-lines
close to the detected emission lines, and will be used to correct for the
intrinsic line widths.

For the X-shooter data reduction we followed the same procedure described in detail in \cite{Kruehler15}, so in the following we will only briefly outline our strategy. First we used the ESO X-shooter pipeline \citep{Goldoni06,Modigliani10} to produce a flat-fielded, rectified and wavelength-calibrated 2D spectrum for every frame in the UVB and VIS arm. For the NIR arm we produced two frames, one for every nod cycle. The resulting 2D frames are then sky-subtracted and rejected for cosmic rays using our own software. The frames are combined by shifting the individual frames and using a weighted average from the signal-to-noise ratio (S/N) to produce one single 2D frame for each arm. We then use a custom-made \texttt{Python} code \citep[again see][]{Kruehler15} to perform optimal extraction on the final 2D frame to obtain the extracted 1D spectrum, using a running Moffat-profile fit to improve the S/N. All wavelengths reported are in vacuum and are corrected for the heliocentric velocity and the Galactic foreground reddening of $E(B-V)=0.082$ mag \citep{Schlafly11}.

\section{Results}    \label{sec:res}

\subsection{Locating the GRB explosion site} \label{subsec:loc}

\begin{figure*}
	\centering
	\epsfig{file=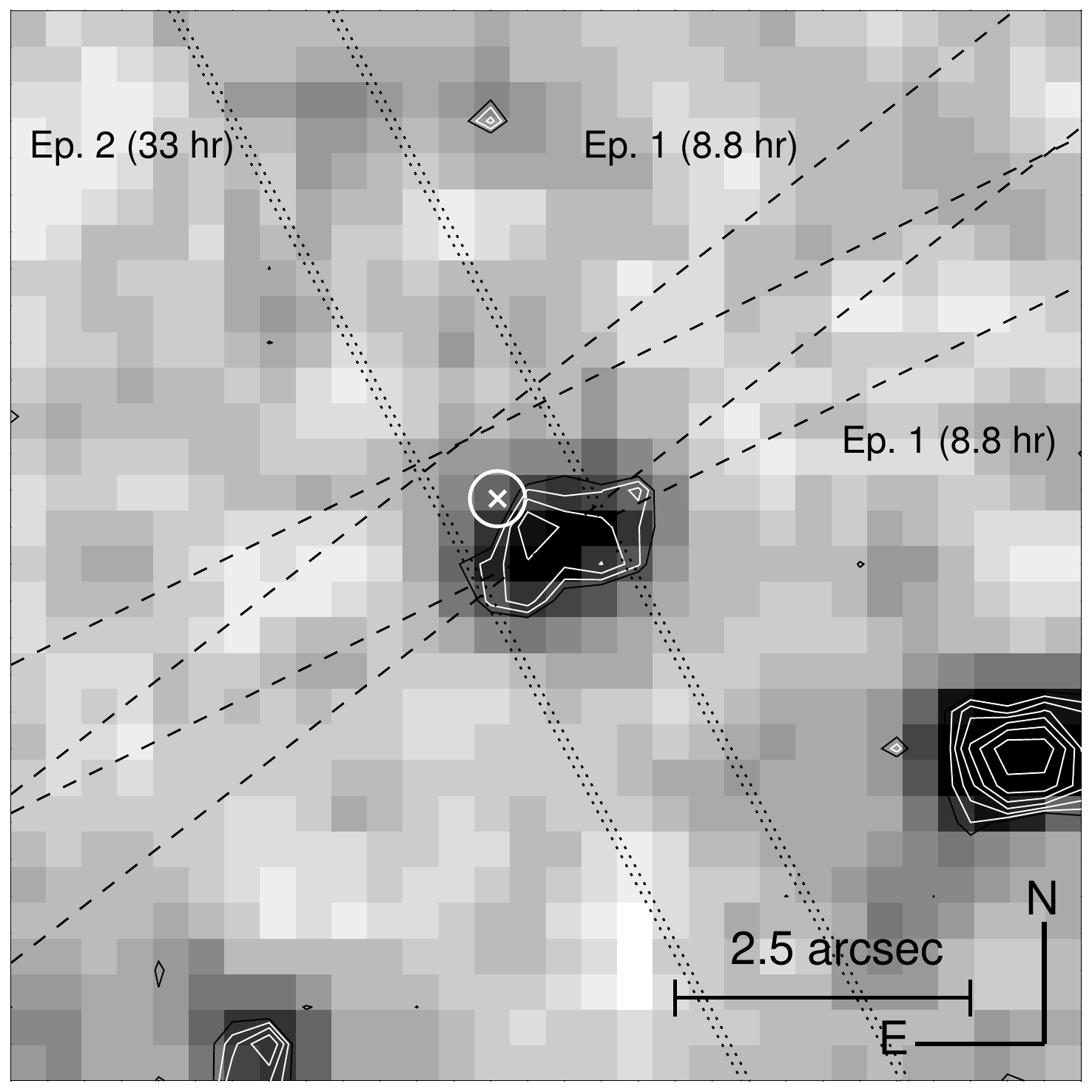,width=8cm}
	\epsfig{file=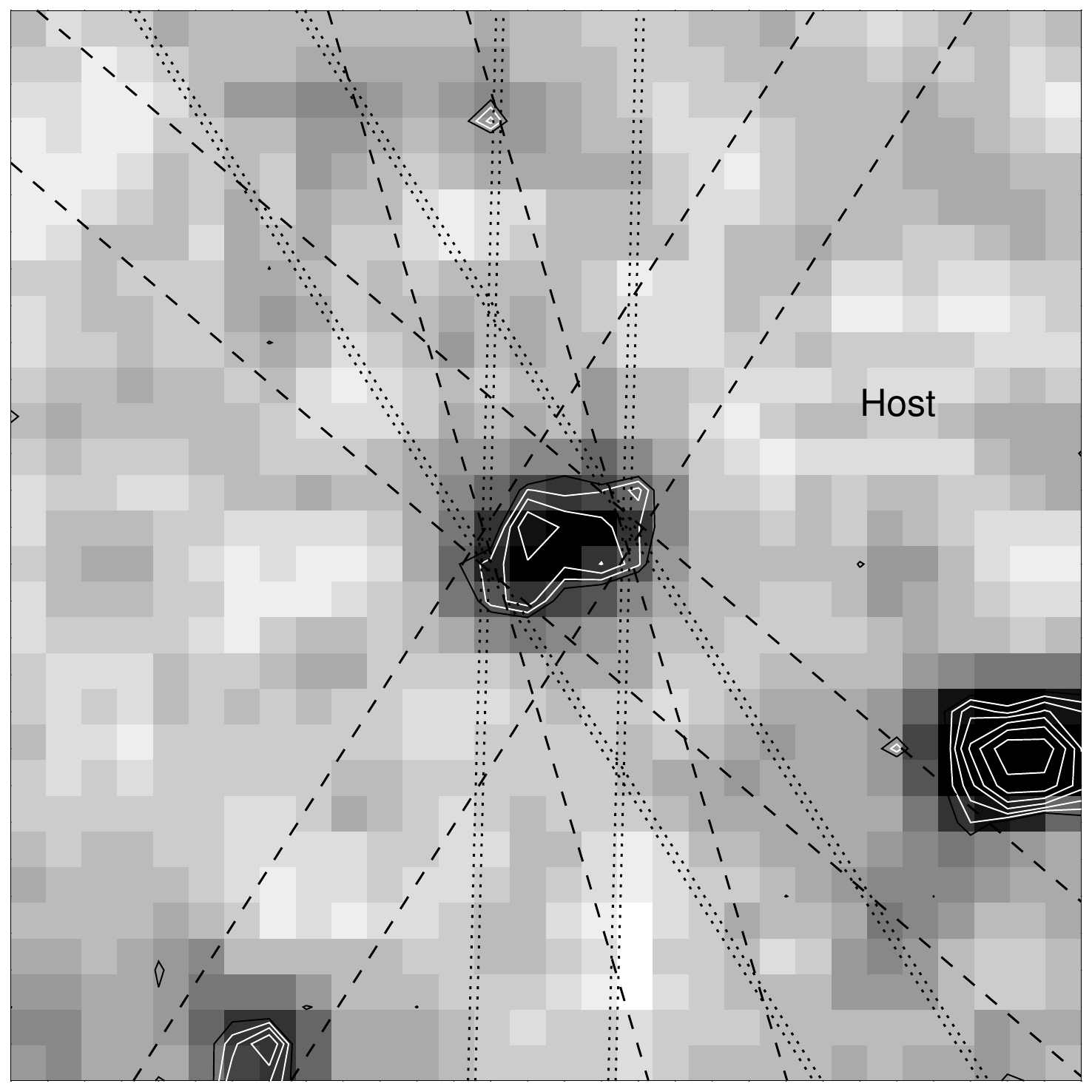,width=8cm}
	\caption{Close-up of the field of GRB\,140506A and its host galaxy from the combined FORS2 $I$-band image ($7.75'' \times 7.75''$ in both panels). The position of the afterglow is offset $0\farcs49\pm 0.03$ ($3.92\pm0.24$ kpc at $z=0.889$) from the host galaxy shown by the white cross with the $3\sigma$ uncertainty on the position shown by the surrounding white circle. Schematics of the slit positions used to observe the afterglow and the host are shown during the first and second epoch (left panel) and from the late-time follow-up (right panel). The image has been smoothed to enhance the contrast and the white contours show the relative image flux levels. \label{fig:grbloc}}
\end{figure*}

An important piece of information concerning the nature of the steep
extinction seen in the optical/near-infrared afterglow of GRB\,140506A is the
location of the explosion site within its host galaxy. In Figure~\ref{fig:grbloc}
we show the projected position of the explosion site relative to the host galaxy in the
stacked $I$-band image obtained with VLT/FORS2. The exact location of the
afterglow is determined by coaligning the host image with an image containing
the afterglow using a set of reference stars common to both images. The respective centres of the afterglow and the host galaxy was determined by fitting a Gaussian profile to the image profiles. The semi-major axis of the host galaxy, derived from the FWHM of the fitted Gaussian profile, is $0\farcs96$ (7.68 kpc at $z = 0.889$). The derived position and its 3$\sigma$ astrometric error is shown by the white cross and white circle, respectively. 

We derive an offset of $0\farcs49\pm 0.03$ ($3.92\pm0.24$ kpc at $z = 0.889$)
from the projected position of the afterglow to the centre of the host and show that it
originated in a faint region in the outskirts of the galaxy \citep[while
GRBs are on average found to explode in the brightest optical regions of their host
galaxies, see e.g.][]{Bloom02,Fruchter06,Lyman17}. We note that higher resolution is required to securely exclude high-surface brightness at the afterglow position that could be washed out due to ground-based seeing. 

The column density of neutral hydrogen ($N_{\mathrm{HI}}$) at the GRB explosion
sites are typically measured from the broad, damped Lyman-$\alpha$ (Ly$\alpha$)
trough for GRBs at $z\gtrsim 1.8$. Due to the low redshift of this system,
however, we are not able detect this absorption feature. From the
\textit{Swift} X-Ray Telescope (XRT) spectra, obtained in photon counting (PC)
mode, an intrinsic column density of $N_{\mathrm{H,X}}=6.7^{+1.4}_{-1.3}\times
10^{21}~\mathrm{cm}^{-2}$ is determined for this burst (see the
\textit{Swift}-XRT GRB spectrum
repository\footnote{\url{http://www.swift.ac.uk/xrt_spectra/}}). This value is
among the highest 10\% of optically detected bursts \citep{Fynbo09}, but is
typical in \textit{Swift} XRT-observed GRB afterglows
\citep[e.g.][]{Campana10,Starling13,Buchner17}. 

\subsection{Excluding the 2175\,\AA~extinction bump scenario}

The extreme flux-deficit below $< 8000$\,\AA~(observer frame) seen in the
spectrum of the afterglow, is suggested by \cite{Fynbo14} to be caused by
either an extreme 2175\,\AA~bump feature in the extinction curve or due to
multiple scattering of light in the nearby environment. The multiple scattering scenario, however, does not seem to be well substantiated in general and was indeed ruled out as the origin of the steep extinction seen towards the supernova SN\,2014J \citep{Johansson17}. The 2175\,\AA~extinction
bump feature was motivated by the fact that the binned afterglow spectrum
appeared to increase in flux below $\sim 4000$\,\AA~which would be predicted if
the absorption feature was indeed due to a bump in the extinction curve. Even
though the 2175 \AA~extinction bump is seen in some GRB afterglows,
they are still rare with only four, robustly detected systems,
GRBs\,070802 \citep{Kruehler08,Eliasdottir09}, 080607
\citep{Prochaska09,Perley11,Zafar11}, 080605 and 080805
\citep{Zafar12}. However, none of these are as extreme as what is seen towards
GRB\,140506A. 

\begin{figure} 
	\centering
	\epsfig{file=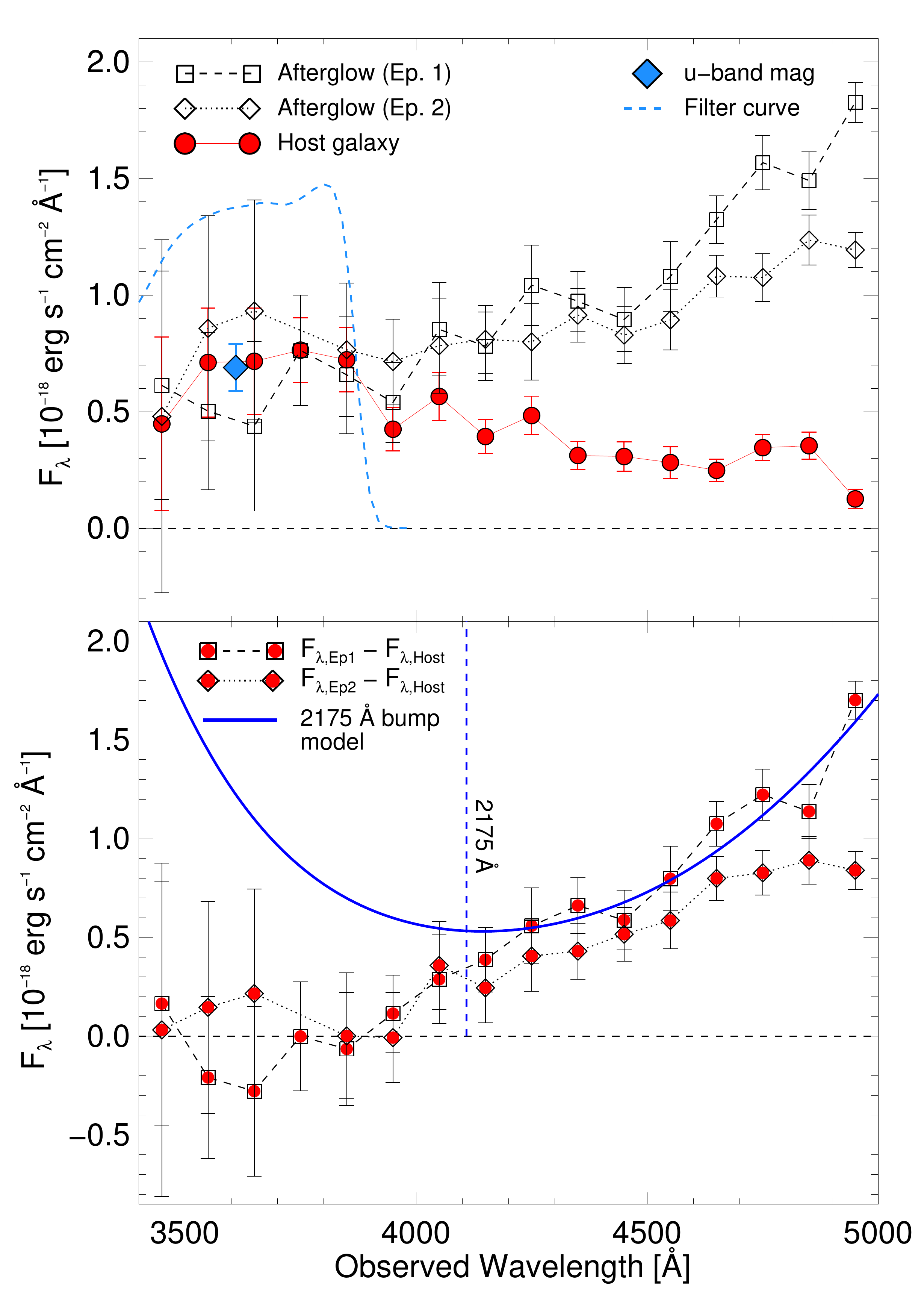,width=9cm,height=10.59cm}
	\caption{(\textit{Top panel}): Comparison of the spectra of the first and second epoch afterglow and the spectrum of the host galaxy at $\lambda < 5000$\,\AA, each binned by 100 \AA, together with the measured VLT/FORS2 $u$-band magnitude of the host. The consistent lower flux in the first epoch afterglow spectrum below 4000\,\AA~is due to the host galaxy being partly outside of the slit (see Fig.~\ref{fig:grbloc}). (\textit{Bottom panel}): First and second epoch afterglow spectra with the flux of the host galaxy subtracted. Overplotted is the 2715\,\AA~extinction bump model from \cite{Fynbo14}, which can now be securely excluded as the scenario causing the steep extinction. \label{fig:hostcontr}}
\end{figure}

To test the hypotheses of the 2175\,\AA~extinction bump, we compare the
contribution of the host galaxy to the afterglow (see
Fig.~\ref{fig:hostcontr}). Following the same procedure as was done for the
afterglow we bin the spectrum of the host galaxy below 5000\,\AA~in 100\,\AA~bins
to match that of the afterglow spectrum. It is evident that the rise seen in
the afterglow spectrum is purely due to the afterglow being host-dominated
below 4000\,\AA~in the first and second epoch. This then indicates that the scenario with the flux-drop being
caused by an extreme 2175 \AA~extinction bump feature can be safely excluded. This is further supported by
the fact that such an extreme bump would be approximately three times as high as seen
toward any line-of-sight in the Milky Way \citep{Fitzpatrick07,Fynbo14}. See Sect.~\ref{ssec:ext} for a parametrization of the extinction curve of the host-corrected first epoch afterglow with the bump or "Drude"-profile removed.

\begin{table}[!ht]
	\centering
	\begin{minipage}{\columnwidth}
		\centering
		\caption{Extracted emission-line fluxes. \label{tab:emlines}}
		\begin{tabular}{lccc}
			\noalign{\smallskip} \hline \hline \noalign{\smallskip}
			Transition & Line flux\tablefootmark{a} & $\sigma$\tablefootmark{b} & $z$ \\ 
			\noalign{\smallskip} \hline \noalign{\smallskip}
			\lbrack O\,\textsc{ii}\rbrack\,$\lambda$$\lambda$\,3726\,,\,3729 & $19.11\pm 0.60$ & $50.8\pm 4.7$ & 0.88905\tablefootmark{c} \\
			H$\beta$ & $3.27\pm 0.73$ & $48.6\pm 15.7$ & 0.88905 \\
			\lbrack O\,\textsc{iii}\rbrack\,$\lambda$\,5007 &  $7.42\pm 0.72$ & $48.0\pm 13.0$ & 0.88905 \\
			H$\alpha$ & $16.47\pm 0.60$ & $49.8 \pm 4.7$ & 0.88905 \\
			\noalign{\smallskip} \hline \noalign{\smallskip}
		\end{tabular}
		\centering
	\end{minipage}
	\tablefoot{
		The values are derived from the best-fit Gaussian functions, measured in a $\pm 30$ \AA~region around the centroid of each line. The listed errors include the propagating errors from the fit.\\
		\tablefoottext{a}{Line fluxes are reported in units of $10^{-18}$ erg cm$^{-2}$ s$^{-1}$.}\\
		\tablefoottext{b}{Velocity dispersions are in units of km s$^{-1}$ and are corrected for the measured resolution in the respective arms.}\\
		\tablefoottext{c}{We only report one redshift and velocity dispersion, since the fit to the [O\,\textsc{ii}] doublet was made with a fixed relation between the two components. }
	}
\end{table}

Fitting the \cite{Fitzpatrick07} extinction curve with the central wavelength of the bump set as a free parameter favors a centre at 2015\,\AA. It is, however, a poor fit and the associated bump width would be a factor of five larger than that typically probed in the Milky Way. This case then leads the way for a non-standard reddening which will have a crucial impact for our understanding of the properties of dust grains in the non-local Universe. We will discuss this more in Sect.~\ref{sec:disc}.

\subsection{Underlying supernova}

The magnitudes of the host galaxy measured in the $R$- and $I$-bands are
approximately one magnitude fainter than what was determined from the GROND late-time photometric follow-up ($R_{\mathrm{FORS2}} - r'_{\mathrm{GROND}} = 0.94\pm0.40$ mag and $I_{\mathrm{FORS2}} - i'_{\mathrm{GROND}} = 0.94\pm0.35$ mag) of the host \citep{Fynbo14}. This could
indicate a contribution from a bright underlying supernova (SN) 70 days ($\approx 35$ days in rest-frame) after trigger during the GROND observations. If this is true, then the GROND magnitudes implies an observed $r'$-band SN luminosty of approximately $1.5\times 10^{39}$ erg s$^{-1}$.

\subsection{Emission line measurements}

We were able to detect the prominent H$\alpha$ and H$\beta$ emission lines as
well as emission from the [O\,\textsc{ii}]\,$\lambda\lambda$\,3726,3729 doublet and the [O\,\textsc{iii}]\,$\lambda$\,5007 transition. To extract the line fluxes we fitted
a Gaussian function to each line with the continuum set in small regions around
the centroid of the fit ($\pm 30$ \AA), free of telluric- and sky-lines. Line
widths were first fixed to a common value in the fits. The fit is dominated by the
strongest line, H$\alpha$. From the fit we then determined the FWHM
for each line. This is converted to a velocity dispersion, $\sigma$, by subtracting the instrumental broadening, $\mathcal{R}$ (km
s$^{-1}$), quadratically from the fitted FWHM as

\begin{equation}
\sigma = \sqrt{\mathrm{FWHM}^2-\mathcal{R}^2}/(2 \sqrt{2\ln 2})~.
\end{equation}

\begin{figure} 
	\centering
	\epsfig{file=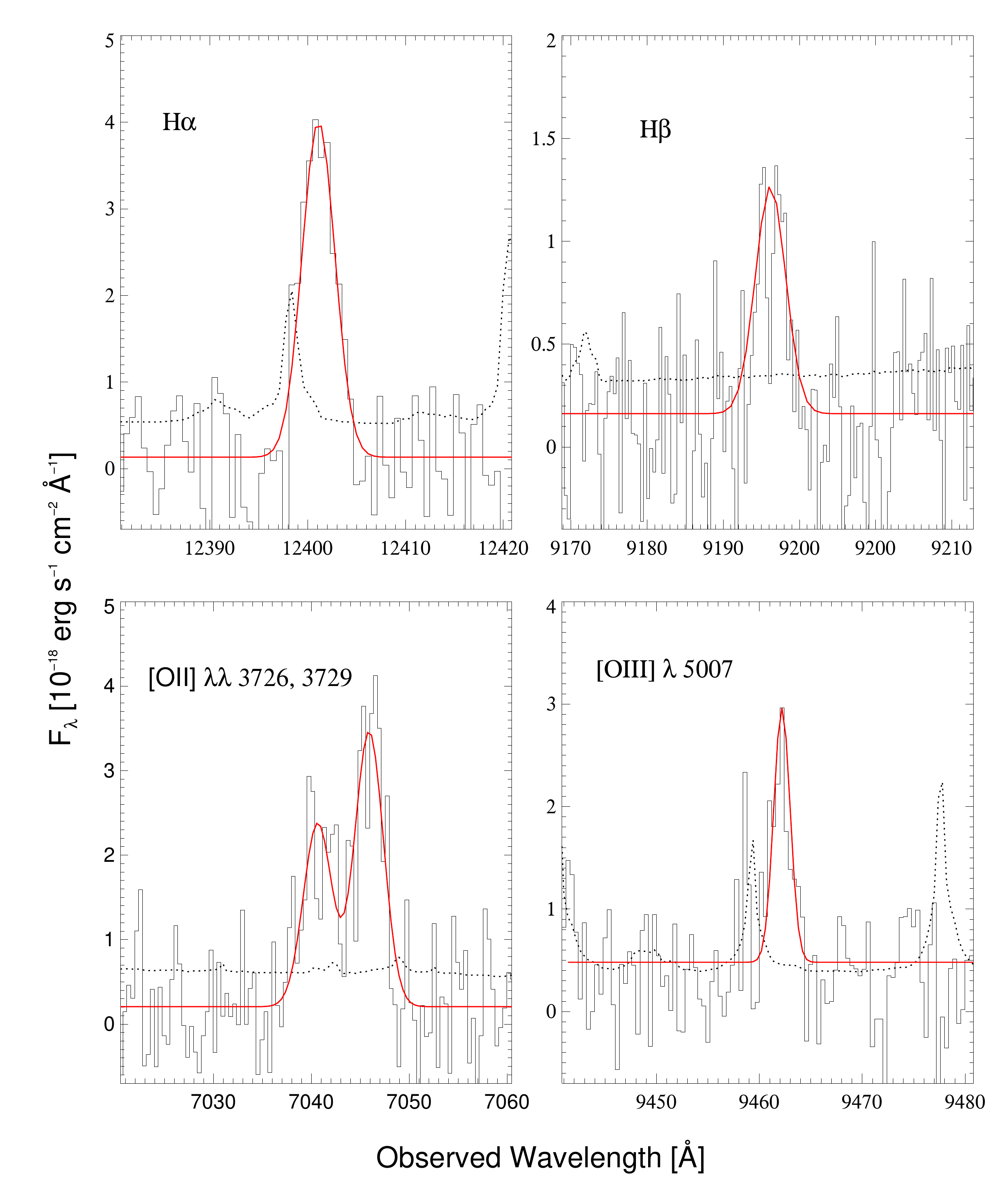,width=9cm}
	\caption{Emission lines from the optimally-extracted 1D spectrum. In each panel the observed spectrum is shown by the solid line, the error spectrum by the dotted line and the best-fit Gaussian function by the solid, red line. The plotting region shows a zoom-in on the continuum region ($\pm 20$ \AA). \label{fig:emlines}}
\end{figure}

Based on the strong H$\alpha$ line we measure an emission-line redshift of
$z=0.88905$ for the host galaxy, consistent with what has been reported in
\cite{Fynbo14} and \cite{Kruehler15}. The results of the measured line fluxes
and line widths are listed in Table~\ref{tab:emlines} and shown in
Fig.~\ref{fig:emlines} with the best-fit Gaussian function.

\subsection{Attenuation of the host} \label{subsec:ext}

The observed H$\alpha$ and H$\beta$ emission line fluxes (the ratio of which is known as the Balmer decrement) provides us with the dust attenuation of the whole system toward the H\,\textsc{ii} regions. The intrinsic, dust-corrected ratio of these two lines, $r_{\mathrm{int}}$, assuming a case B recombination and an electron temperature of $10^4$ K and a density of $10^2 - 10^4$ cm$^{-3}$, is H$\alpha$/H$\beta=2.87$ \citep{Osterbrock89}. Using the MW-type extinction curve of \cite{Pei92} with $R_V=3.08$, we find that the attenuation at the wavelength of H$\alpha$ and H$\beta$ is $A_{\mathrm{H}\alpha}=0.82$ mag and $A_{\mathrm{H}\beta}=1.17$ mag, for this specific extinction curve normalised to $A_V=1$ mag, respectively. We thus calculate the visual attenuation toward the H\,\textsc{ii} regions from the Balmer decrement as
\begin{equation}
A_V=\frac{-2.5\log(r_{\mathrm{obs}}/r_{\mathrm{int}})}{A_{\mathrm{H}\alpha}-A_{\mathrm{H}\beta}} = 1.74\pm 0.41~\mathrm{mag}~,
\end{equation}
where in our case, $r_{\mathrm{obs}}$ is found to be
$(\mathrm{H}\alpha/\mathrm{H}\beta)_{\mathrm{obs}}=5.04\pm 0.67$. We note that
modelling the spectrum with an SMC-like extinction curve instead \citep[also
typically seen in GRB host galaxies;][]{Schady10} will not change the results
significantly since there is little difference at the wavelength range of the
Balmer lines, and that most extinction curves behave similar redwards of the
2175\,\AA~extinction bump. The MW extinction curve of \cite{Pei92} was also
chosen to be consistent with \cite{Kruehler15}. See Sect.~\ref{sec:disc} for
further discussion on the reddening of the host and the comparison to other GRB
host galaxies.

\subsection{Star formation rate}

To measure the star formation rate (SFR) of the host galaxy we calculated the intrinsic, dust-corrected luminosity of the H$\alpha$ emission line using the luminosity distance computed from the redshift, $z=0.88905$. This yields a de-reddened line luminosity of $L(\mathrm{H}\alpha)=(2.61\pm 0.09)\times 10^{41}$ erg s$^{-1}$, where the applied dust-correction was calculated as
\begin{equation} \label{eq:dustcorr}
L_{\mathrm{dered}}=L_{\mathrm{int}}\times 10^{(0.4\,\times\,A_{\lambda})}~ .
\end{equation}
Using the relation from \cite{Kennicutt98} and assuming a Salpeter initial mass function (IMF) we calculate a dust-corrected SFR of
\begin{equation}
\mathrm{SFR(H}\alpha) = 7.9\,\times 10^{-42}\,L(\mathrm{H}\alpha)_{\mathrm{dered}}=2.14\pm 0.07~M_{\odot}~\mathrm{yr}^{-1}~,
\end{equation}
where, assuming a Chabrier IMF instead, this can be converted to
$\mathrm{SFR(H}\alpha)_{\mathrm{Chab}}=1.34\pm 0.04~M_{\odot}~\mathrm{yr}^{-1}$
\citep{Treyer07}. For consistency with \cite{Kruehler15} we refer only to the latter value in Sect.~\ref{sec:disc} in our comparison.

\subsection{Metallicity} \label{subsec:meta}

To infer the emission-line metallicity of the host galaxy it is common practice
to use the strong-line ratios; [N\,\textsc{ii}]/[O\,\textsc{ii}] (N2O2), ([O\,\textsc{ii}]\,$\lambda$\,3727+[O\,\textsc{iii}]\,$\lambda$\,4959+[O\,\textsc{iii}]\,$\lambda$\,5007)/H$\beta$ ($R_{23}$), [N\,\textsc{ii}]/H$\alpha$
(N2) and ([O\,\textsc{ii}/H$\beta$)/([N\,\textsc{ii}]/H$\alpha$) (O3N2). A
detailed discussion on how the strong-line diagnostics can be used as
metallicity indicators and a comparison of these is given in \cite{Kewley08}.
In the following we assume a solar metallicity of $12+\log(\mathrm{O/H})=8.69$
\citep{Asplund09}. 

\begin{figure} 
	\centering
	\epsfig{file=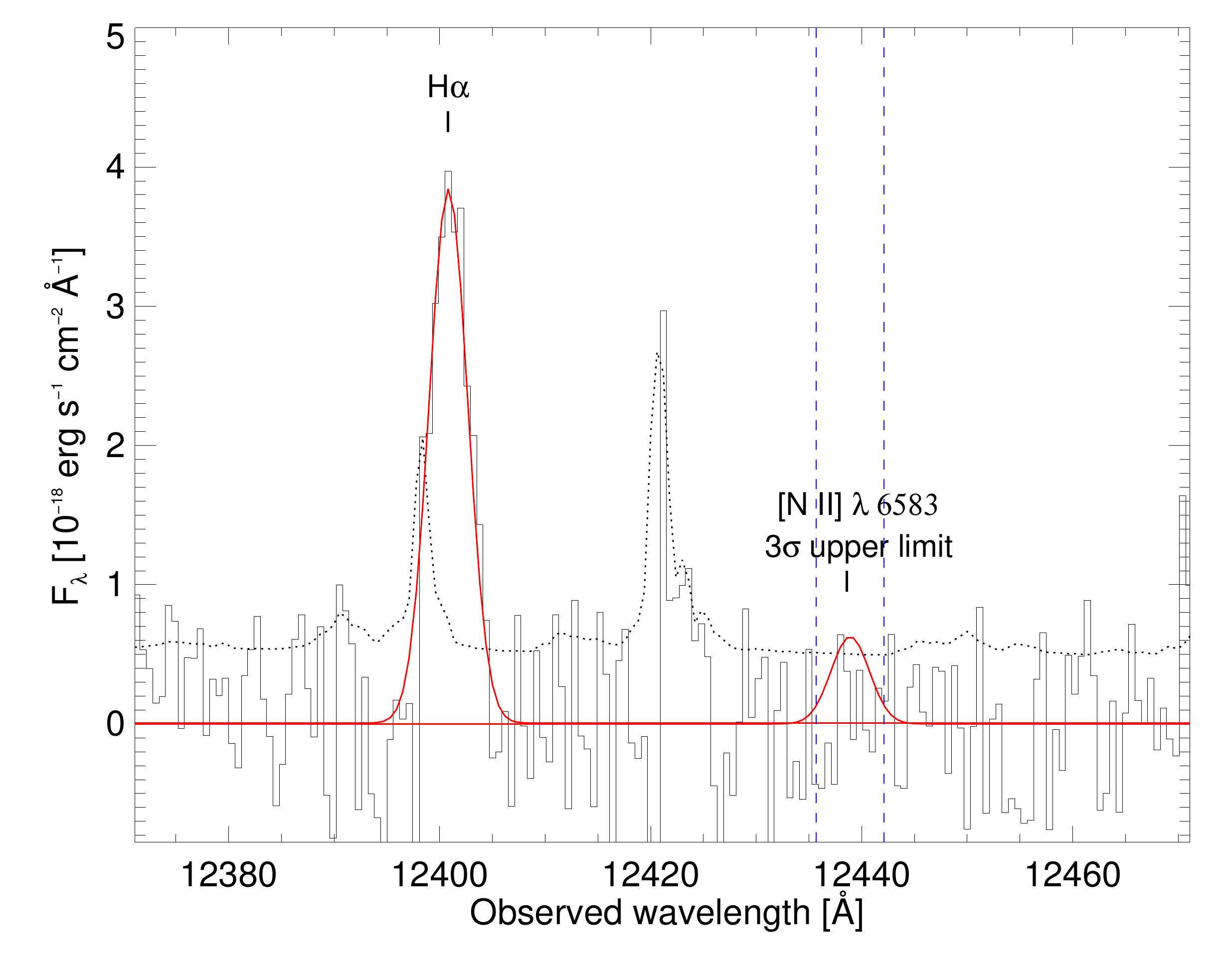,width=9cm}
	\caption{The $3\sigma$ flux limit of [N\,\textsc{ii}] measured from the continuum compared to H$\alpha$. We assumed a Gaussian shape of the line profile and measured the limit at the FWHM (dashed blue lines) of that of H$\alpha$. The total flux of the continuum at [N\,\textsc{ii}] was then computed from the flux encompassed in this region. \label{fig:N2upperlim}}
\end{figure}

Due to the non-detection of [N\,\textsc{ii}] in our case
(Fig.~\ref{fig:N2upperlim}), we are only able to determine upper limits from
the aforementioned strong-line ratios from the $3\sigma$ upper limit measured
from the continuum at the wavelength of [N\,\textsc{ii}]. From the
dust-corrected emission-line fluxes (Eq.~\ref{eq:dustcorr}) we compute upper limits on the oxygen
abundances of $12+\log(\mathrm{O/H})<8.46$ for N2O2
\citep{Kewley02}, $12+\log(\mathrm{O/H}) < 8.37$ for N2
\citep{Pettini04} and $12+\log(\mathrm{O/H}) < 8.67$ for
O3N2 \citep{Pettini04} corresponding to upper limits of $0.59, 0.49~\&~0.95~Z/Z_{\odot} $, respectively. The errors and upper limits on all derivations include
the scatter in the relations listed in \cite{Kewley08}.

\subsubsection{$f(P,R_{23})$ calibration}

To alleviate the limitations from the non-detection of [N\,\textsc{ii}] we make use of the fact that it is also possible to estimate the metallicity based solely on the strong oxygen nebular lines \citep{Pilyugin01,Pilyugin05}, the so-called $P$ method. The authors advocate that the physical properties of the H\,\textsc{ii} region can be estimated via the excitation parameter, $P$, here defined as $P=R_3/(R_2+R_3)$, where $R_2=$[O\,\textsc{ii}]/H$\beta$ and $R_3=$([O\,\textsc{iii}]\,$\lambda$\,4959+[O\,\textsc{iii}]\,$\lambda$\,5007)/H$\beta$. Since we do not detect [O\,\textsc{iii}]\,$\lambda$\,4959 we use that $f_{[\mathrm{O}\,\textsc{iii}]\,\lambda\,4959}=1/3\times f_{[\mathrm{O}\,\textsc{iii}]\,\lambda\,5007}$ \citep{Storey00}.

Following the above definitions, we derive an excitation parameter of $P=0.21\pm0.03$. To compute the metallicity we use the relation 
\begin{equation}
12+\log(\mathrm{O/H})=\frac{R_3+106.4P+106.8P^2-3.40P^3}{17.72P+6.60P^2+6.95P^3-0.302R_3}~,
\end{equation}
from \cite{Pilyugin05}. Using the value for the excitation parameter found above and the ratio $R_3=2.81\pm0.45$ this yields an oxygen abundance of $12+\log(\mathrm{O/H})=8.23\pm0.16$, corresponding to 35\% solar metallicity ($Z = 0.35^{+0.15}_{-0.11}~Z_{\odot}$). The expected scatter of 0.1 dex in the relation is included in the calculation. This is consistent with the previous upper limits. We will only refer to this value in the remainder of this paper.

\section{Discussion} \label{sec:disc}

\subsection{Broad-band SED of the dusty host} \label{ssec:sed}

\begin{figure} 
	\centering
	\epsfig{file=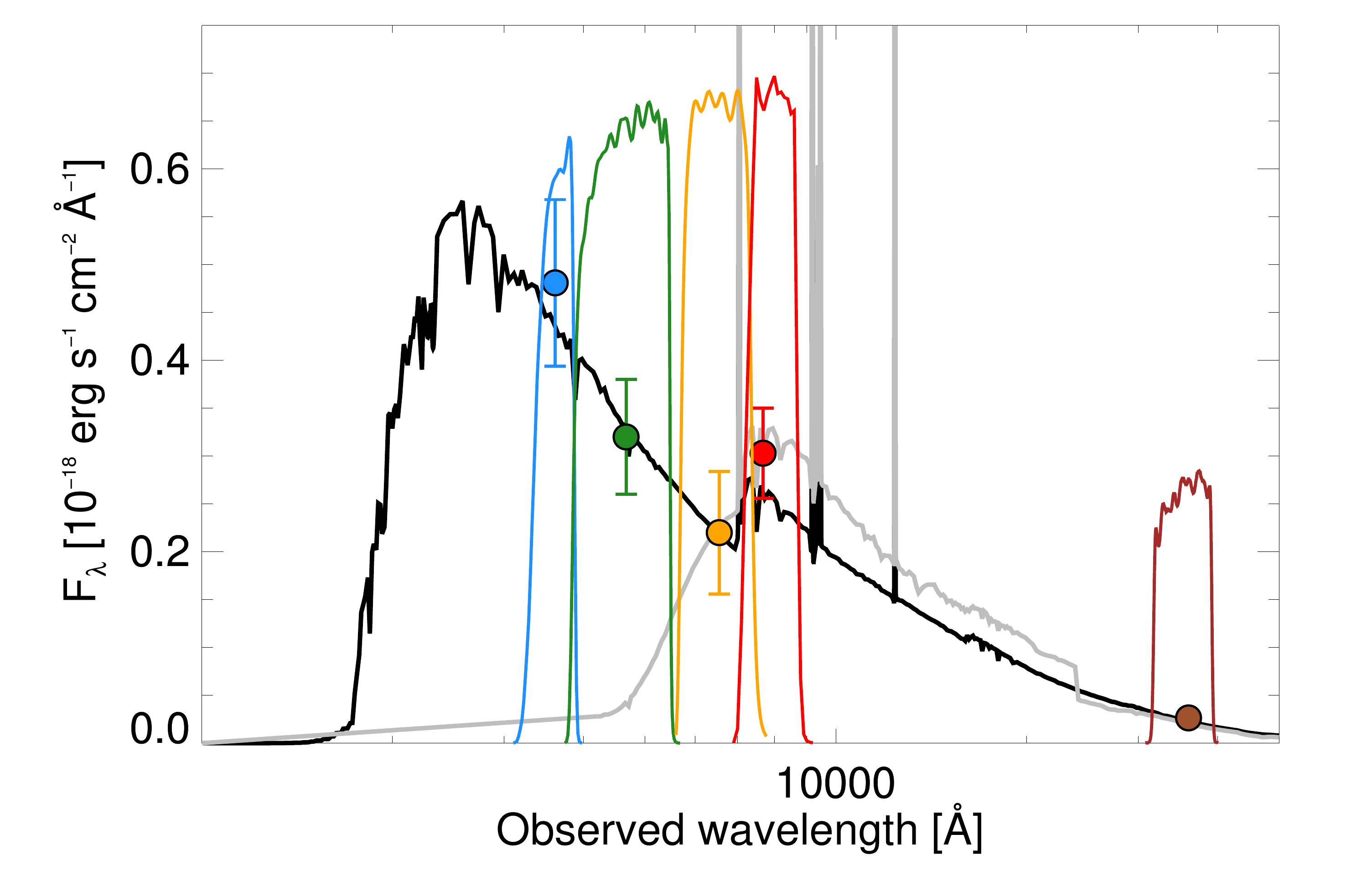,width=9cm}
	\caption{SED of the host galaxy of GRB\,140506A. Overplotted is the
		best-fit stellar population synthesis model of a galaxy with a 20 Myr old stellar population at a fixed
		redshift of $z=0.88905$, reddened by $A_V=1.62$ mag (black line) following the extinction
		curve of \cite{Calzetti00}. Overplotted is the same intrinsic SED but reddened by the steep extinction curve seen towards the afterglow instead (grey line) but without the rise in flux at wavelengths below 4000\,\AA~(see text). Also shown are the filter transmission curves for
		the corresponding five photometric points. The data do not suggest an atypical
		steep extinction curve for the host, similar to the afterglow sightline.
		\label{fig:hostsed}}
\end{figure}

To constrain the physical origin of the steep extinction curve seen towards the sightline of GRB\,140506A, we first model the SED of the host galaxy. Fitting the observed five broad-band
magnitudes reported in Table~\ref{tab:photdata} in
LePhare\footnote{\url{http://www.cfht.hawaii.edu/~arnouts/LEPHARE/lephare.html}} \citep{Arnouts99,Ilbert06} yields the host galaxy parameters listed in
Table~\ref{tab:sedfit}, with a $\chi^2=0.98$, and the best-fit stellar population synthesis model
shown in Fig.~\ref{fig:hostsed}. We obtained the best fit SED by fixing the
redshift to $z=0.88905$ and used a grid of stellar evolution models with
varying star formation time scales, age of stellar population and extinction
assuming the models from \cite{Bruzual03} based on an IMF from \cite{Chabrier03} and a Calzetti extinction curve \citep{Calzetti00}.

The best-fit visual attenuation, $A_V^{SED}$, and star-formation rate from the broad-band SED agree well with that determined from the emission-line diagnostics. The stellar mass computed by the fit is mainly dominated by the 3.6 $\mu$m \textit{Spitzer} photometric data point.

It is evident that the host galaxy does not follow an extinction curve as steep as that seen towards the GRB explosion site. Running the SED fit with the new extinction curve (see next section) in steps of $E(B-V)=[0.0,\,0.05,\, ...\,, 1.0]$ we find a best-fit with no attenuation, that is all models with the steep extinction curve implemented provides worse fit to the data. As a conclusion, the steep extinction curve toward the afterglow must be sightline dependent and this is evidence for the occurrance of only locally steep extinction curves. The dust properties of the intermediate environment in which the GRB is embedded, strongly imprinted on the afterglow SED, are therefore not representative of the dust properties of the host galaxy as a whole. 

\begin{table}
	\caption{Host galaxy parameters from SED fitting.}
	\label{tab:sedfit}
	\centering
	\begin{tabular}{lllr}
		\noalign{\smallskip} \hline \hline \noalign{\smallskip}
		Extinction law & & & \cite{Calzetti00} \\
		Reddening $A_V$ [mag] & & & $1.62\pm0.21$ \\
		SFR [$M_{\odot}$ yr$^{-1}$] & & & $1.18\pm0.28$ \\
		Age [Gyr] & & & $0.02^{+0.07}_{-0.02}$ \\
		$\log(M_*/M_{\odot})$ & & & $8.35 \pm 0.26$ \\
		\noalign{\smallskip} \hline \hline \noalign{\smallskip}
	\end{tabular}
\end{table}

\subsection{Modelling the extinction curve} \label{ssec:ext}

To further confine the shape of the extinction curve we subtract
the flux of the host galaxy from the best-fit SED from the first epoch
afterglow. To estimate the reddening we assume that the intrinsic spectrum
follows a power-law, $F_{\lambda}\propto \lambda^{-\beta}$, with $\beta = 1.75$
as derived from the XRT spectrum with a break between the X-rays and optical
bands of $\Delta\beta = 0.5$ \citep[following the same procedure
as][]{Fynbo14}. We then run a non-linear least squares curve fitting on the
host-subtracted afterglow spectrum using \texttt{MPFIT} in \texttt{IDL} with
the parametrization of \cite{Fitzpatrick07}, which models the extinction curve through a set of nine parameters. It basically contains two components, one describing the UV linear part of the extinction specified by the parameters $c_1$ (intercept) and $c_2$ (slope) and the parameters $c_4$ and $c_5$ provide the far-UV curvature and the other is the Drude component describing the 2175\,\AA~bump by the parameters $c_3$ (bump strength), $x_0$ (central wave number) and $\gamma$ (width of the bump). The last two parameters are the visual extinction, $A_V$, and the total-to-selective reddening, or the steepness of the reddening law, $R_V$.

\begin{figure*} [ht!]
	\centering
	\epsfig{file=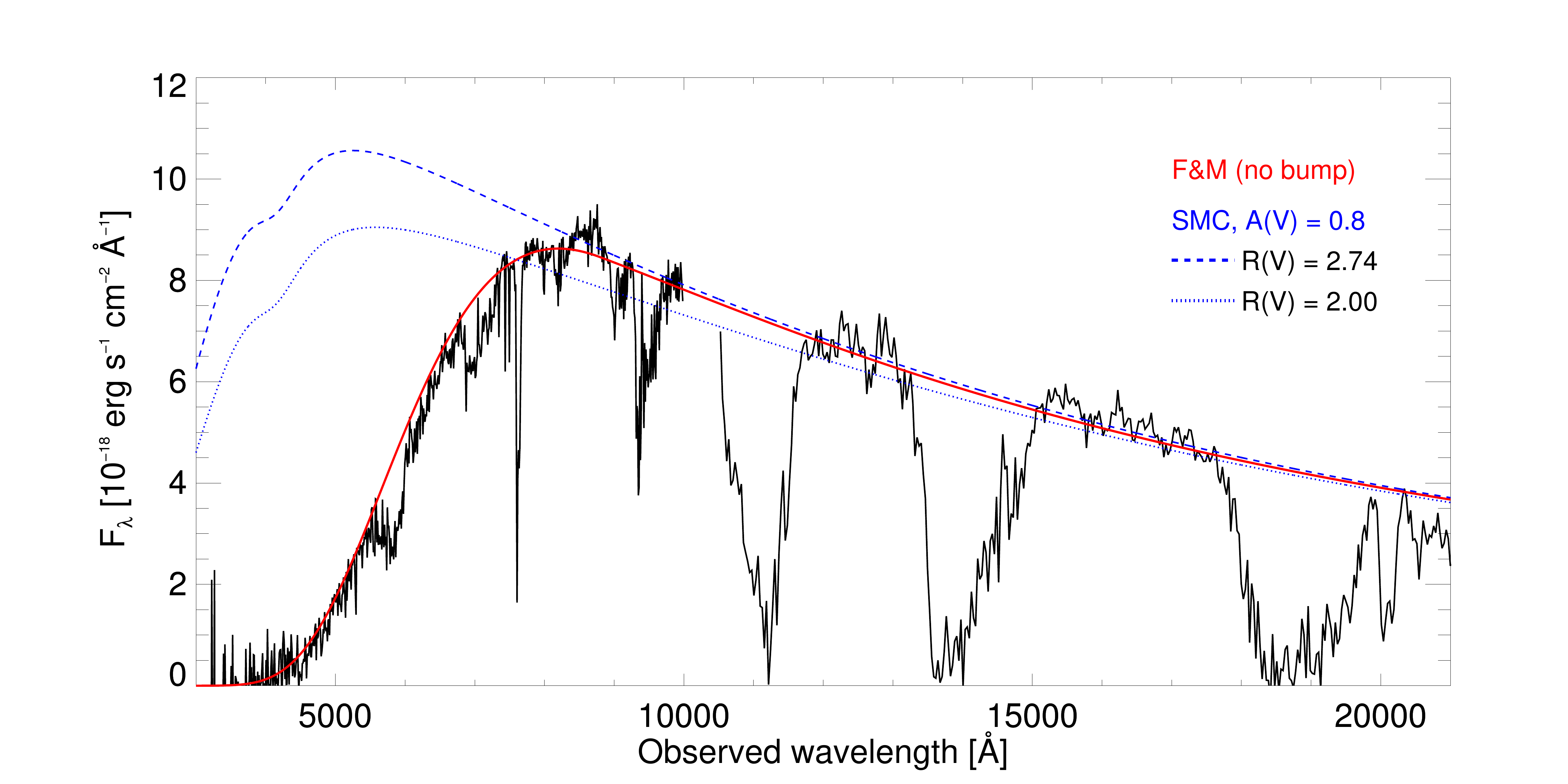,width=16.5cm}
	\caption{First epoch afterglow spectra with the flux of the host galaxy from the best-fit SED subtracted. Overplotted is the best-fit \cite{Fitzpatrick07} parametrization of the extinction curve with $c_3=0$ (no bump) shown by the red solid line. We assume an underlying power-law with slope set by a fit to the X-ray afterglow and assuming a $\Delta\beta = 0.5$ cooling break between the X-rays and the optical. Regions affected by telluric absorption are removed from the fit. Also plotted is the extinction curve of the SMC as described by \cite{Gordon03} (blue dashed line) and with an decreased total-to-selective reddening, $R_V$ (blue dotted line). These examples proves to show that the extinction can not simply be modelled by a standard or even steeper SMC-type reddening law. The parameters of the best-fit extinction curve diverges significantly from that known from the local analogs, most notable are the $c_4$ and $c_5$ values representing the far-UV curvature component, but also the flat value of $R_V$. These are required to model the extreme drop without assuming a bump in the extinction curve in combination with the otherwise blue spectrum.
		\label{fig:ext}}
\end{figure*}

Setting all nine parameters free we find a best-fit almost identical to
the extreme 2175\,\AA~extinction bump proposed by \cite{Fynbo14}. However,
having excluded any signs of a bump in the extinction curve we instead run the
fit assuming $c_3=0$, effectively removing the Drude profile representing the
bump in the model. We then find the best-fit values of: $c_1=-0.93$,
$c_2=3.13$, $c_4=3.30$, $c_5=2.15$, $R_V=5.12$ and $A_V=1.04$ mag, resulting in the
extinction curve shown in Fig.~\ref{fig:ext} and is defined as
\begin{multline}
\frac{A_{\lambda}}{A_V} = \frac{1}{5.12} \left(-0.93+3.13x\right)+1~\mathrm{for}~x<c_5 ~~~\& \\
= \frac{1}{5.12} \left(-0.93+3.13x+3.30(x-2.15)^2 \right)+1~\mathrm{for}~x>c_5~, 
\end{multline}
where $x=\lambda^{-1}$ in units of $\mu$m$^{-1}$.
Most notable are the flat or grey
total-to-selective extintion parameter, $R_V$, and the parameter describing the
far-UV curvature component, $c_4$. Comparing these to a simple SMC-type
extinction curve as modelled by \cite{Gordon03} reveals that a typical steeper
reddening law, commonly defined by a small value of $R_V$, is inadequate in reproducing the strong drop. Instead a
flat extinction curve is required, mainly to reproduce the relatively blue part
of the spectrum at wavelengths above 8000\,\AA, with the roughly seven times
larger value of $c_4$ compared to that of the SMC \citep{Gordon03} modelling
the steep drop. The parameters describing the linear component of the UV extinction, $c_1$ and $c_2$, also differentiate significantly compared to the relation $c_1=2.09 - 2.84 c_2$ found by \cite{Fitzpatrick07}, where we would expext $c_1=-6.80$ (best-fit $c_1=-0.93$) based on the $c_2$ parameter.

\begin{figure} [h!]
	\centering
	\epsfig{file=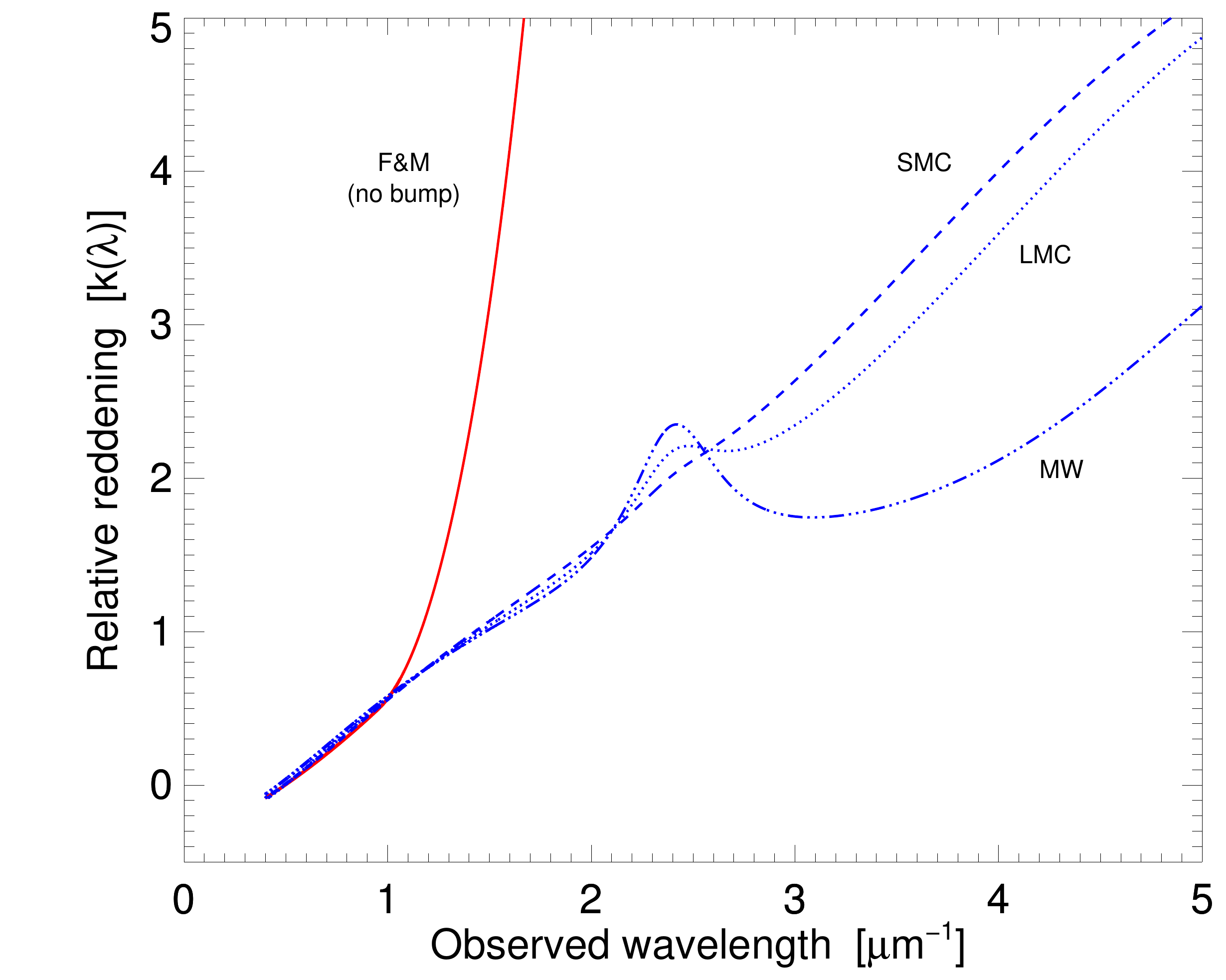,width=8.8cm,height=8.2cm}
	\caption{Reddening curves for the first epoch afterglow spectra with the flux of the host galaxy from the best-fit SED subtracted. Since we do not know the instrinsic brightness of the afterglow the data has been normalised to the flux in the $K$-band ($\sim 20\,000$\,\AA). The reddening curves, described by $k(\lambda)$, thus only indicate the relative reddening and not the total extinction. It is clearly shown how the extinction curves of the Local Group, here the MW, SMC and LMC, can only match the best-fit reddening law from the \cite{Fitzpatrick07} parametrization at wavelengths longer than 8000\,\AA~(observers frame), at which point the steep extinction curve diverges significantly.
		\label{fig:red}}
\end{figure}

From the above derived parametrization of the extinction we attempt to
recover the relative reddening curve as well. By again assuming the XRT spectrum with a
cooling break as the instrinsic afterglow spectrum, we normalise this power-law
to the $K$-band (at roughly 20\,000\,\AA~in regions with no telluric
absorption) and then we calculate the relative reddening, $k(\lambda)$, as
\begin{equation}
k(\lambda) = -2.5\log\left(D_{\lambda}/I_{\lambda}\right)~,
\end{equation} 
where $D_{\lambda}$ denotes the observed data and $I_{\lambda}$ refers to
the intrinsic spectrum. The resulting reddening curve is shown in
Fig.~\ref{fig:red} where we also compare it to the standard extinction curves
of the Local Group (that of the MW, the SMC and the LMC). It is evident
that the standard reddening laws do not provide a good fit to the data,
except in the reddest part of the spectrum.

\subsection{Comparison to GRB hosts at $z<1$}

\begin{figure*} 
	\centering
	\epsfig{file=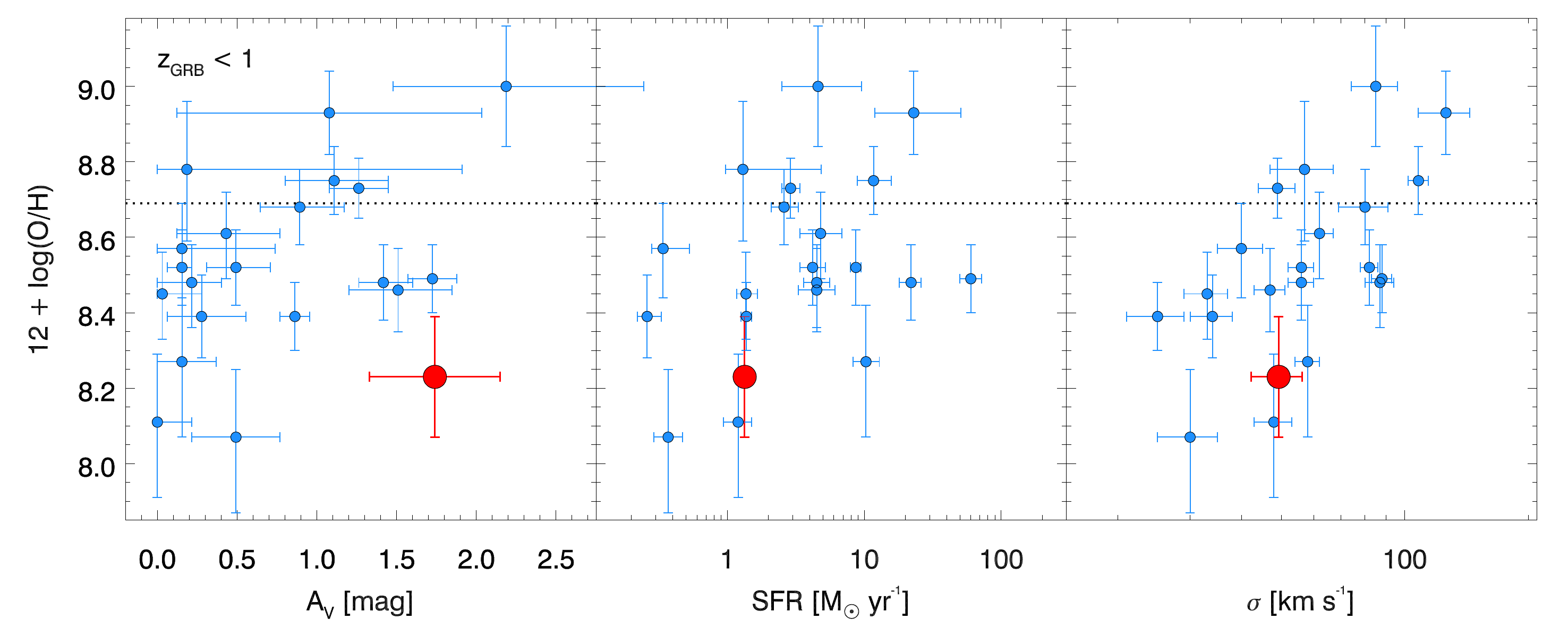,width=17cm}
	\caption{Comparison of the host galaxy properties of GRB\,140506A (large red dot) to 20 other GRB host galaxies at $z<1$ (small blue dots) with metallicity measurements from \cite{Kruehler15}. From left to right the metallicity of each individual host galaxy is shown as a function of: $A_V$, SFR and $\sigma$, measured from emission line diagnostics. The dotted, horizontal line denotes the solar metallicity abundunce \citep{Asplund09}. The physical properties of the host galaxy of GRB\,140506A is inconspicuous, compared to the other GRB host galaxies at $z<1$. 
		\label{fig:hostprop}}
\end{figure*}

Another potential clue to understand why the afterglow of GRB\,140506A is so different from the typical SED of GRB afterglows, is to compare the properties of the host galaxy to other GRB hosts at similar redshifts. We use a subsample of 20 GRB host galaxies at $z<1$ with metallicity measurements extracted from the full sample of 96 GRB-selected galaxies presented in \cite{Kruehler15}. We caution that this is not an unbiased sample but merely a compilation of GRB host galaxies observed with X-shooter. In Fig.~\ref{fig:hostprop} we show the physical properties of this subsample and compare these GRB host galaxies with GRB\,140506A. 
It is evident that this particular host galaxy is a standard GRB host in terms of star formation rate, velocity dispersion and metallicity but with a large, though not abnormal, visual attenuation, $A_V$. 

\subsection{Comparison to SNe and AGNs}

One of the main differences between observed GRB and SN samples is that the explosion sites of GRBs are found much closer to the brighter regions of their respective host galaxies \citep{Fruchter06,Lyman17}. 

Numerous more cases of SNe with steep extinction curves are observed \citep[e.g.][]{Krisciunas06,Nobili08,Folatelli10}, which are argued could imply a grain size that is small in the specific line of sights compared with the average value for the local ISM \citep{Elias-Rosa06,Elias-Rosa08} or a dominating multiple scattering process \citep{Goobar08}. See also \cite{Amanullah14} for a prime example of a steep extinction curve seen towards SN 2014J in M82. Again we wish to highlight that in this specific case the multiple scattering scenario was ruled out \citep{Johansson17}. \cite{Hutton14} were able to constrain the global reddening law of M82, showing that standard MW-type extinction is preferred for the galaxy as a whole, which is further evidence that steep extinction is only seen locally. In these studies it is argued that the smaller grain size causing the steep extinction could be fragmentations of larger dust grains disolved due to the large radiation field. 

A similar steep extinction curve is observed toward several dust-reddened quasars as well. \cite{Fynbo13} found that $\sim 30\%$ of the quasars in their sample were reddened following an extinction curve much steeper than what is seen in the local analogs; MW, SMC or LMC \citep[see also][for a specific example towards the AGN Mrk 231]{Leighly14}. A similar type of extinction is also seen (but to a lesser extent) in the High $A_V$ Quasar \citep[HAQ;][]{Krogager15} and the \textit{extended} HAQ \citep[eHAQ;][]{Krogager16} surveys. A subsample of 16 intrinsically reddened quasars from the HAQ survey were examined by \cite{Zafar15}, who found a weighted mean value of $R_V=2.4$, much steeper than that of the SMC which has a total-to-selective extinction ratio around $R_V\sim 2.9$. We note that even these steep extinction curves do not provide a good fit to the data, due to the still close resemblance to that of the SMC.

\subsection{Variability of the flux drop}

The above cases further suggests that the steep extinction curve observed
towards GRB\,140506A is not a rare phenomenon, but is more likely attributed to
the level of fragmentation of the dust grains surrounding the explosion site.
More curios is the temporal variability of the first to the second epoch
afterglow spectrum. From a study of the dynamics of large dust grains in
molecular clouds, \cite{Hopkins16} show that grains with sizes $a>0.01$ $\mu$m
can exhibit large fluctuations in the local dust-to-gas ratio. They conclude
that small dust grains can be much more clumped than larger dust grains,
critical for dust growth and shattering. 

If that is the case for the circumburst medium then this might explain the temporally varying afterglow SED. Due to
the decrease of the relativistic beaming as the blast wave slows down, the
emitting area of the afterglow expands as ($\gamma c t$)$^2$ \citep[see also the discussion in][]{Fynbo14}, which yields a factor of six increase in the glowing
region producing the afterglow between 8.8 and 33 hr post-burst. If the probed
distribution of small and large dust grains is different for such different
beam sizes, then this will have an effect on the observed extinction towards
the line of sight to the GRB at the two different epochs.  

\section{Conclusions} \label{sec:conc}

We have analysed the host galaxy of the mysterious GRB\,140506A afterglow, showing an extreme flux-drop below 8000\,\AA~observers frame (4000\,\AA~rest-frame) in the sightline towards the GRB explosion site \citep{Fynbo14}. This is atypical to the blue power-law SEDs commonly observed for GRB afterglows \citep[e.g.][]{Fynbo09}. We found that the explosion site of the GRB occurred at a projected distance of $\approx$ 4 kpc from the galactic centre, in a faint region in the outskirts of the host galaxy, while most GRBs are observed much closer to the centre or in the brightest regions of their respective host galaxies \citep{Fruchter06,Lyman17}. 

We found the host galaxy to be moderately forming stars at a rate of $1.34\pm 0.04~M_{\odot}$ yr$^{-1}$.
Employing the $P$ method we derived a metallicity of $12+\log(\mathrm{O/H})=8.23\pm0.16$ ($Z = 0.35^{+0.15}_{-0.11}~Z_{\odot}$). This is in good agreement with the $Z-\mathrm{SFR}$ and $Z-\sigma$ relations found for GRB host galaxies at $z<1$ \citep{Kruehler15}. 
The host galaxy can be characterised by a large visual attenuation of $A_V=1.74\pm 0.41$ mag based on emission line diagnostics. 

We excluded the scenario where the rise in flux observed at $<4000$\,\AA~in the afterglow spectrum was due to an extreme 2175\,\AA~extinction bump, a feature known from the local LMC and the MW, and showed that it was merely a representation of the dominating host galaxy light at this wavelength. There does not seem to be any similarity between the extinction law of the afterglow and the best-fit for the host galaxy which suggests that the steep extinction seen in the afterglow SED is sightline dependent and only a local effect. The $A_V$'s measured in both cases are high, with a best-fit afterglow extinction of $A_V=1.04$ mag. This effect could be explained by the intense radiation field from the GRB only fragmenting (in contrast to fully destroying) the surrounding dust particles causing a local only steep extinction.

We modelled the relative reddening and showed that all the standard extinction curves as seen in the Local Group (MW, SMC and LMC) are inadequate in descring the steep drop in combination with the relatively blue underlying power-law spectrum.
A similar non-standard, steep extinction curve is observed toward a few other rare cases of GRBs but also in several other types of enigmatic astrophysical objects, such as AGNs and SNe. The phenomenon of steep extinction curves does not seem to be rare globally and also hint that this could contribute (at least partly) to the large fraction of dark GRBs, defined by no optically detected afterglows. 
A new picture of the dust properties in high energy environment seems to be emerging. The extinction curves with non-local analogs appear to be an important clue to resolve this.

\begin{acknowledgements}
We would like to thank the anonymous referee whose comments greatly improved the quality of the paper and arrived in a timely manner. 
We would also like to thank C. Gall, D. Malesani and B. T. Draine for insightful discussions on the conclusions of the paper. KEH and PJ acknowledge support by a Project Grant (162948--051) from The Icelandic Research Fund. The research leading to these results has received funding from the European Research Council under the European Union's Seventh Framework Program (FP7/2007--2013)/ERC Grant agreement no.  EGGS--278202. TK acknowledges support through the Sofja Kovalevskaja Award to P. Schady from the Alexander von Humboldt Foundation of Germany. LC is supported by YDUN grant DFF -- 4090--00079. D.X. acknowledges the support by the One-Hundred-Talent Program of the Chinese Academy of Sciences (CAS), by the Strategic Priority Research Program “Multi-wavelength Gravitational Wave Universe” of the CAS (No. XDB23000000), and by the National Natural Science Foundation of China under grant 11533003.
\end{acknowledgements}

\bibliographystyle{aa}
\bibliography{ref}

\end{document}